\documentstyle[preprint,aps]{revtex}
\input epsf.sty
\tighten

\begin{document}
\draft

\title{\Large \bf Corrections to Scaling for the Two-dimensional 
                 Dynamic  XY Model}

\author{\bf  H.P. Ying$^{1,2}$, B. Zheng$^{1,3}$,
                         Y. Yu$^{2}$ and S. Trimper$^{3}$}

\address{$^1$Zhejiang Institute of Modern Physics,
          Zhejiang University, Hangzhou 310027, P.R. China}
\address{$^2$Institute of Theoretical Physics, Academia Sinica,
            100080 Beijing, P.R. China}
\address{$^3$FB Physik, Universit\"at Halle, D--06099 Halle, Germany}

\maketitle

\begin{abstract}
With large-scale Monte Carlo simulations, we confirm that
for the two-dimensional XY model, there is 
a logarithmic correction to scaling in the dynamic relaxation
starting from a completely disordered state, while 
only an inverse power law correction in the case of
starting from an ordered state. The dynamic exponent $z$
is $z=2.04(1)$.

\end{abstract}


In recent years, much attention has been drawn to non-equilibrium
short-time behavior of critical dynamics.
Traditionally, it is believed that universal dynamic scaling
behavior only exists in the long-time regime of the dynamic evolution.
In 1989, however, with renormalization group methods
 Janssen, Schaub and Schmittmann
derived a dynamic scaling form for the $O(N)$ vector model,
which is valid up to the {\it macroscopic} short-time regime
\cite {jan89}.
The dynamic process they considered is that the system initially
at a very high temperature state with small or zero magnetization,
 is suddenly quenched to the critical
temperature, then released to dynamic evolution of model A. 
Important is that a new independent critical exponent
must be introduced to describe the scaling behavior of
the initial magnetization. Afterwards, 
some evidences for the short-time dynamic scaling
were also observed in Monte Carlo simulations
\cite {hus89,hum91}.
On the other hand, it was found that 
the power law decay of the magnetization starting
from a completely ordered state shows up
from rather early times, e.g., see \cite {sta92,ito93},
and it can be used
to estimate the dynamic exponent $z$. 
Inspired by these works, in the last several years
non-equilibrium short-time critical dynamics
has been systematically investigated
with Monte Carlo methods 
\cite {li94,sch95,gra95,zhe98,zhe99}.
Simulations have extended from regular classical spin models
\cite {zhe98,zhe99b,bra00,jen00,san00},
to statistical systems with quenched disorder
\cite {luo98,luo99,yin00}, quantum spin systems
and lattice gauge theories \cite {yin98,oka98,san00a}, 
dynamic systems without detailed balance 
\cite {men98,tom98a,bru99},
the hard-disk model \cite {jas99a,jas00} and fluid systems
\cite {zha00}. 
References given here are only a part of recent ones
and not complete. A relatively complete list of
the relevant references before 1998 can be found
in Ref. \cite {zhe98}.
All numerical and analytical results confirm 
the existence of a rather general dynamic scaling form
in critical dynamic systems at early times.

The short-time dynamic scaling has not been systematically
explored in experiments. But 
the dynamic scaling behavior around a spin-glass transition
\cite {hus89,fis91,blu92,kis96,luo99}
is very similar to that around a standard critical point.
For example, the experimental measurements of the remanent
magnetization in spin glasses support not only the power law scaling behavior
but also the scaling relations between the 
exponents \cite {gra87,luo99}.

The short-time dynamic scaling form not only is 
conceptually interesting, but also provides
new techniques for the measurements of
both dynamic and static critical exponents as well as
the critical temperature \cite {sta92,blu92,li95,gra95},
 for a review see Ref. \cite {zhe98}.
Since now the measurements are carried out
in the short-time regime, the dynamic approach does not suffer 
from critical slowing down.
Compared with those methods developed in equilibrium,
e.g., the non-local cluster algorithms,
the dynamic approach does study the original local
dynamics and can be applied to disorder systems.
Furthermore, to solve numerically dynamic equations with a continuous time
to the long-time regime is very difficult,
but the short-time dynamic approach works well
\cite {zhe99}. Such a method should be also very interesting
in experiments.

To understand the universal short-time
behavior, one should distinguish 
the macroscopic and microscopic time scales.
The dynamic scaling emerges only after
a time scale $t_{mic}$ which is sufficiently large
in microscopic sense but still very small in macroscopic sense.
In Monte Carlo simulations, for example, if a sweep over
all spins on a lattice is considered to be
a microscopic time unit,
$t_{mic}$ is usually from a few to 100 Monte Carlo time steps
\cite {zhe98}. Therefore, performing simulations up to
some hundred or thousand time steps is usually sufficient
to obtain rather good values for critical exponents.
However, in the recent study of the two-dimensional XY model
(with a Kosterlitz-Thouless phase transition)
 and the random-bond Ising model
\cite {bra00,luo98b,bra00a,luo00},
one observes somewhat unexpected phenomena.
The dynamic exponent $z$ estimated from a dynamic
processes starting from a disordered state is bigger  
(10 to 15 percent for the XY model and 
5 to 10 percent for the random-bond Ising model) than that from
an ordered state.
Puzzling is that the resulting static exponents
are correct within statistical errors. 
This behavior should have its origin in the existence
of the free vortices or the metastable states.
Similar concern for the XY model with different 
boundary conditions and dynamics can be found also
in Ref. \cite {jen00}.
If such a kind of phenomena are not clarified,
further applications of the short-time dynamic scaling
becomes complicated and difficult.

In a recent Letter \cite {bra00} (see also Ref. \cite {bra00a}),
Bray, Briant and Jervis argue theoretically 
that there is a logarithmic correction 
for the two-dimensional XY model 
in the dynamic process
starting from a disordered state. However,
the presented numerical data can not distinguish the
two ansatzes, a possible bigger $z$ or a logarithmic correction.
On the other hand, there has been some controversy over
the value of the dynamic exponent $z$ (see, e.g.,
Ref. \cite {jen00} and references therein), 
and here it is especially interesting
whether $z$ is
exactly 2 \cite {bra00}.
In this Communication, we report our large-scale Monte Carlo simulations
for the two-dimensional XY model,
examine possible corrections to scaling in dynamic processes
starting from both ordered and disordered states
and determine relevant critical exponents accurately.

In the simulations, the system in a macroscopic initial
state, is suddenly quenched to the transition temperature
$T_{KT}$ or slightly below, 
then released to dynamic evolution of model A.
In the literature,
$T_{KT}$ is reported to be between
$0.89$ and $0.90$. In this paper, we take the temperature $T=0.89$.
 Following Ref. \cite {bra00},
we adopt the 'heat-bath' algorithm in which a trial move
is accepted with probability $1/[1+exp(\Delta E/T)]$,
where $\Delta E$ is the energy change associated with the move.
All results are presented with a lattice size $L=256$.
Simulations with other lattice sizes confirm that the finite 
size effect for $L=256$ has been completely invisible
for our updating times. 

Denoting a spin $S_i(t)$, as usual, we define
the magnetization, its second moment and
the auto-correlation as  
$\vec M(t) \equiv \langle \sum_i \vec S_i(t) \rangle /L^2$,
$M^{(2)}(t) \equiv \langle [ \sum_i \vec S_i(t) ]^2 \rangle /L^4$
and $A(t) \equiv \langle \sum_i \vec S_i(0) \cdot \vec S_i(t) \rangle /L^2$
respectively.

\begin{figure}[t]\centering 
\epsfysize=10.cm 
\epsfclipoff 
\fboxsep=0pt
\setlength{\unitlength}{1cm} 
\begin{picture}(13.6,12.)(0,0)
\put(-1.,0){{\epsffile{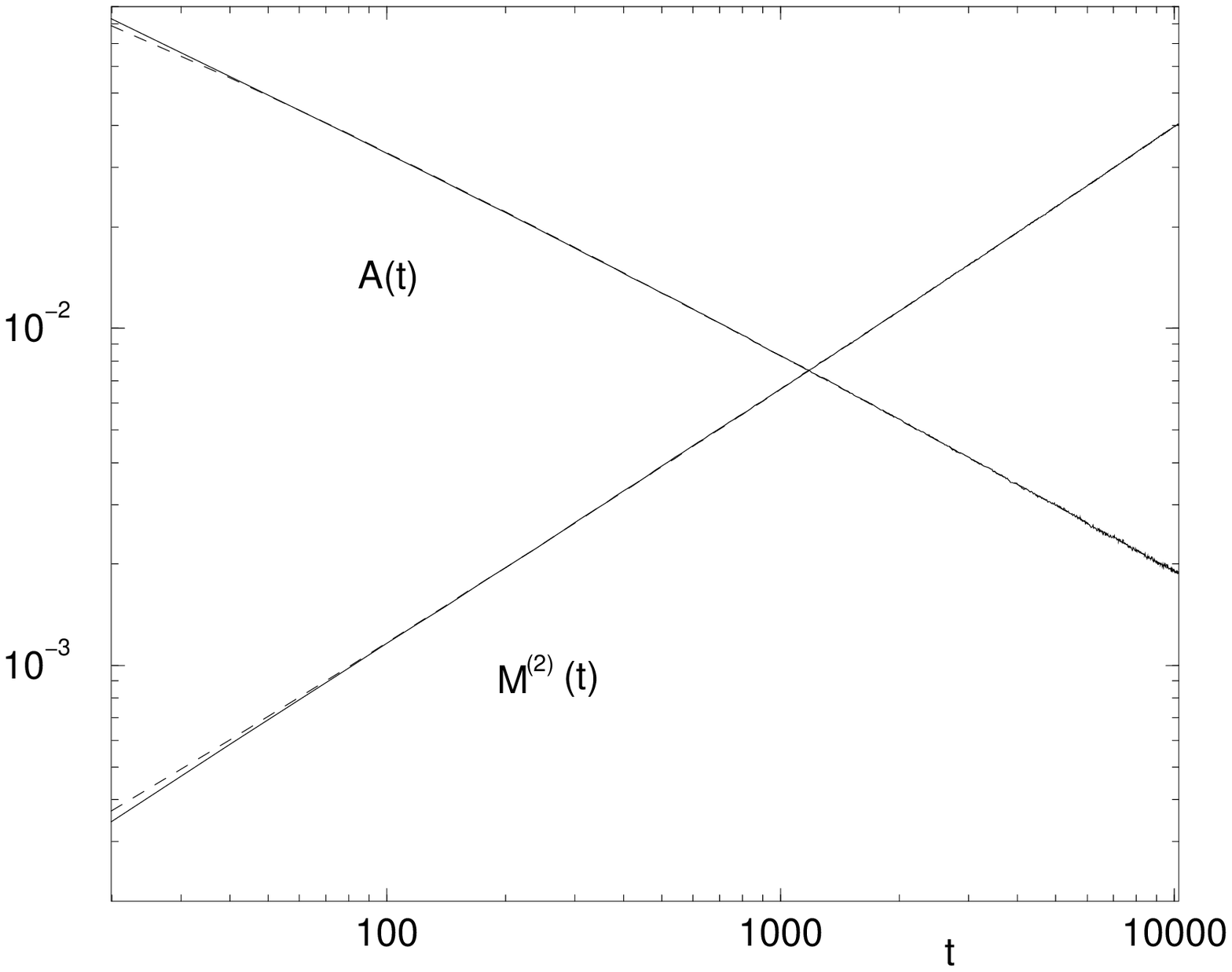}}} 
\end{picture} 
\caption{Time evolution of the second moment
and auto-correlation starting from a disordered state
in log-log scale. Dashes lines are the fitted curves
with a logarithmic correction.
} 
\label{f1}
\end{figure}

In Fig. \ref {f1}, time evolution of
the second moment and the auto-correlation
for a {\it disordered} initial state are displayed with solid lines
 in log-log scale.
In order to detect any corrections
to scaling, we have performed the simulations up to
$t=10240$ Monte Carlo time steps.
 Samples of the initial configurations (also random numbers)
for averaging are $20000$. To estimate the errors,
samples are divided into four subsamples.
Assuming that there is a logarithmic correction
for the non-equilibrium spatial correlation length,
according to general scaling analysis,
the second moment should behave like \cite {bra00}
\begin{equation}
M^{(2)}(t) = b_2 [t/(1+c_2 ln (t))]^{(2-\eta)/z} \ .
\label{e1}
\end{equation}
Here $\eta$ is the usual static exponent, $z$ is the dynamic
exponent and $b_2$ and $c_2$ are constants.
Similarly, the auto-correlation
\begin{equation}
A(t) = b_a [t/(1+c_a ln (t))]^{\theta-d/z} \ .
\label{e2}
\end{equation}
Here $d=2$ is the spatial dimension.
If $c_2$ and $c_a$ are zeros, the standard power law scaling
behavior is recovered. Looking at Fig. \ref {f1},
$A(t)$ bends obviously downwards, in consistence with
the logarithmic correction. However, the behavior of $M^{(2)}(t)$
is somewhat complicated and the correction is also
less strong than that for $A(t)$.
 It bends slightly downwards at early times,
and changes to upwards only after about 100 time steps.
The first behavior is not universal behavior
but microscopic details dependent. Anyway,
if the simulation is performed only up to
$t=2000$ or $3000$ \cite {luo98b,bra00},
 it would be difficult to conclude
whether and how the power law is corrected.
Now, we fit the two solid lines in Fig. \ref {f1} to
the ansatzes in Eqs. (\ref {e1}) and (\ref {e2})
in a time interval $[100, 10240]$.
The fitted curves are shown with dashed lines
in the figure.  The quality of the fitting is good
and the resulting exponents are
$(2-\eta)/z=0.866(3)$ and $d/z-\theta=0.730(1)$.

Here it is very important to address that
 if directly measuring the
slope, e.g., for $M^{(2)}(t)$ in Fig. \ref {f1},
{\it in any time intervals} we obtain $(2-\eta)/z$ around $0.77$ to
$0.78$. These values differ from $0.866(3)$ 
by more than 10 percent. Do $M^{(2)}(t)$ and 
$A(t)$ fit {\it uniquely} to the ansatzes 
in Eqs. (\ref {e1}) and (\ref {e2})?
We have tried inverse power law corrections, e.g., 
for $M^{(2)}(t)$, 
\begin{equation}
M^{(2)}(t) \sim t^{(2-\eta)/z}(1+c/t^b)\ .
\label{e3}
\end{equation}
For both $M^{(2)}(t)$ and $A(t)$, the quality of the fitting
is even slightly better than with a logarithmic correction.
However, the correction exponent $b$ is small,
$b=0.211$ and $0.0474$ for $M^{(2)}(t)$ and $A(t)$
respectively, while the exponent $d/z-\theta$
remains the same and $(2-\eta)/z$ differs only 
by 1 or 2 percent. This strongly indicates that
a logarithmic correction is indeed correct.
It is believed that the logarithmic corrections 
are related to the vortex pair annihilation, 
and do not disappear 
within a time scale $t_{mic}$ \cite {bra00,bra00a}.

\begin{figure}[t]\centering 
\epsfysize=10.cm 
\epsfclipoff 
\fboxsep=0pt
\setlength{\unitlength}{1cm} 
\begin{picture}(13.6,12.)(0,0)
\put(-1.,0){{\epsffile{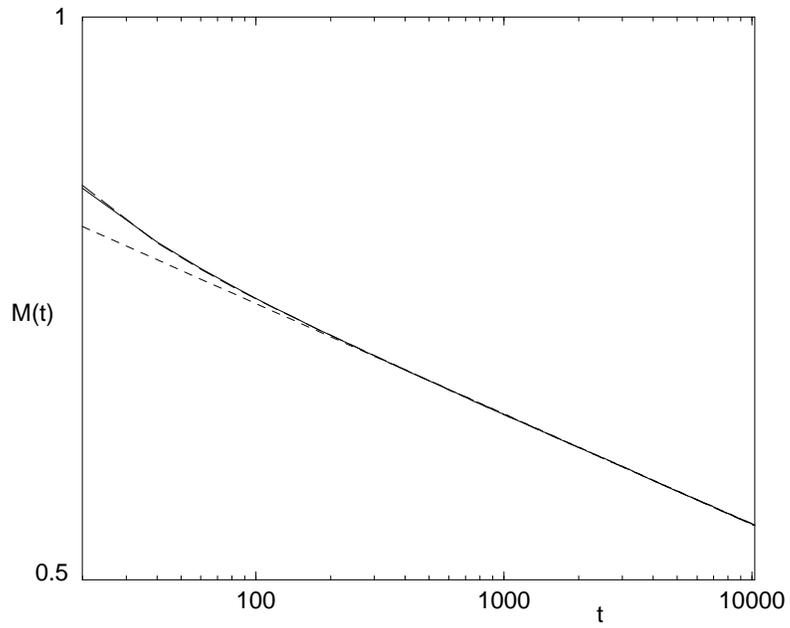}}} 
\end{picture} 
\caption{Time evolution of the magnetization starting from
an ordered state in log-log scale. The dashed line is
for a power law fit and the long dashed line is with 
an inverse power law correction.
} 
\label{f2}
\end{figure}

\begin{figure}[t]\centering 
\epsfysize=10.cm 
\epsfclipoff 
\fboxsep=0pt
\setlength{\unitlength}{1cm} 
\begin{picture}(13.6,12.)(0,0)
\put(-1.,0){{\epsffile{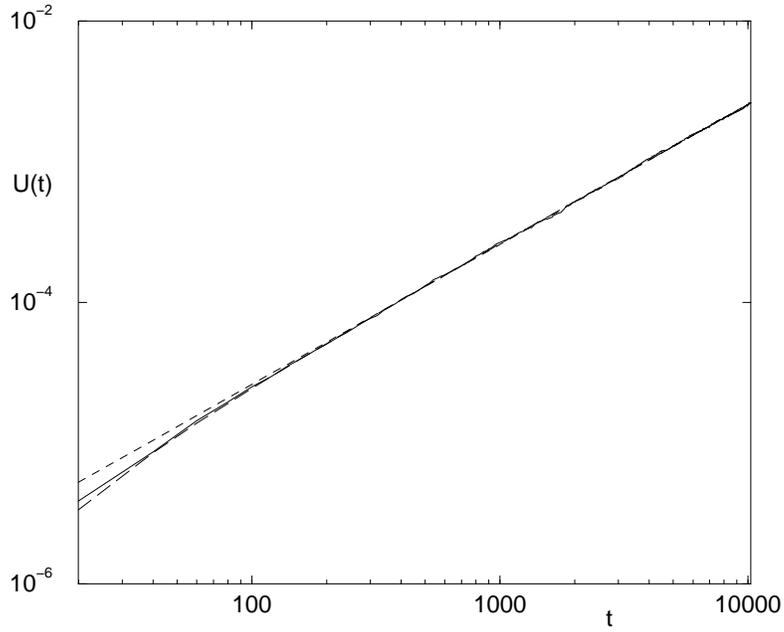}}} 
\end{picture} 
\caption{Time evolution of the Binder cumulant starting from
an ordered state in log-log scale. The dashed line is
for a power law fit and the long dashed line is with 
an inverse power law correction.
} 
\label{f3}
\end{figure}

For a dynamic process starting from an ordered state,
i.e. $\vec M(0)=(1,0)$, no logarithmic corrections 
are claimed theoretically,
since no free vortices exist.
It is interesting to confirm this numerically and obtain independently
the dynamic exponent $z$ and the static exponent $\eta$
for comparison. In this dynamic process,
the magnetization $\vec M(t)=(M(t),0)$ 
is subject to the power law scaling 
behavior \cite {zhe98}
\begin{equation}
M(t) \sim t^{-\eta/2 z}\ .
\label{e4}
\end{equation}
In order to determine the dynamic exponent $z$ independently,
we introduce a time-dependent Binder cumulant,
$U \equiv M^{(2)}/M^2-1$, which behaves like
\begin{equation}
U(t) \sim t^{d/z}\ .
\label{e5}
\end{equation}
In Fig. \ref {f2} and Fig. \ref {f3},
$M(t)$ and $U(t)$ are displayed with solid lines
in log-log scale. Samples (now only respect to random numbers)
for averaging are $10000$. Both curves show deviation
from power law up to $t \sim 200$ or $300$.
However, a logarithmic correction does not fit
to the curves. Therefore, 
we should either accept a relatively bigger
$t_{mic}$, or consider inverse power law corrections.
With an ansatz similar to Eq. (\ref {e3}), in a time interval
$[100, 10240]$ we obtain
$\eta/2 z = 0.0588(3)$ and $d/z = 0.982(10)$.
The fitted curves are shown with long-dashed lines
in Fig. \ref {f2} and Fig. \ref {f3}. They overlap nicely with
the numerical data (solid lines).
Without considering corrections to scaling, the estimated exponents
differ about 1 percent
(with relatively bigger $t_{mic}$. The corresponding curves
are shown with dashed lines in Fig. \ref {f2} and Fig. \ref {f3}.

\begin{figure}[t]\centering 
\epsfysize=10.cm 
\epsfclipoff 
\fboxsep=0pt
\setlength{\unitlength}{1cm} 
\begin{picture}(13.6,12.)(0,0)
\put(-1.,0){{\epsffile{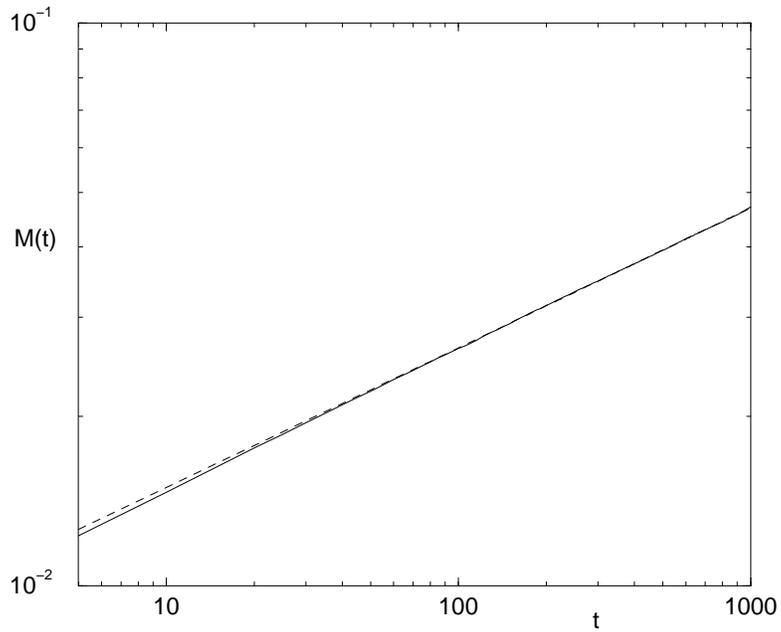}}} 
\end{picture} 
\caption{Time evolution of the magnetization starting from
a disordered state but with a small initial value 
in log-log scale. The dashed line is
for a power law fit.
} 
\label{f4}
\end{figure}

Finally, to complete our investigation we study
a dynamic process starting from a disordered state
but with a {\it small} initial magnetization $\vec M(0)=(m_0,0)$.
If assuming a dynamic scaling form, one can deduce
that at the early times, the magnetization 
$\vec M(t)=(M(t),0)$ obeys 
a power law \cite {zhe98}
\begin{equation}
M(t) \sim t^{\theta}\ .
\label{e6}
\end{equation}
Here $\theta$ is a new independent critical exponent
related to the initial condition \cite {jan89,zhe98}.
Since we need a small initial magnetization $m_0$
and suffer from large fluctuation in longer times,
the simulation is only performed up to $t=1000$. 
samples for averaging is $14000$. In Fig. \ref {f4},
$M(t)$ is displayed with a solid line on log-log
scale. From these data, we can not detect a logarithmic 
correction. In a time interval $[100, 1000]$,
direct measurement of the slope yields an exponent 
$\theta=0.250(2)$, which is the same as 
considering an inverse power law correction.
The dashed line in Fig. \ref {f4} corresponds to
a simple power law fit.
Of course, we can not exclude that a logarithmic correction
may be detected if we perform simulations up to
$t=10000$. But data analysis of the exponents below will 
show that this will very probably not happen.

In Table \ref {t1}, we summarize all the measured exponents.
For the dynamic process starting from an ordered state,
through the measured $d/z$ we can obtain 
independently the dynamic exponent $z$, denoted as $z_1$ in the table.
Then, with $z_1$ as input,
we calculate the static exponent $\eta=0.240(3)$ from
 $\eta/2 z$. This value is slightly bigger than $\eta=0.23$
estimated in simulations in equilibrium \cite {gup92},
but we believe our value is more accurate.
With  $\eta$ in hand, from the index $(2-\eta)/z$ 
in the dynamic process starting from a disordered state,
we estimate another value $z_2=2.03(1)$ for the dynamic exponent $z$.
Finally, combining the results of
 $\theta$ and $d/z-\theta$ we obtain the third value
$z_3=2.04(1)$. Three estimates of $z$ from different 
dynamic processes agree very well.
This supports the logarithmic corrections
in Eqs. (\ref {e1}) and (\ref {e2}). A remark here is that
even if there might be a logarithmic correction
for the magnetization in Eq. (\ref {e6}), it must be rather weak
and $\theta$ would not be
modified so much, otherwise $z_3$ will deviate from
$z_1$ and $z_2$.
Our impression is that even a small initial magnetization
would suppress the effect of the vortex pairs.

Without considering a logarithmic correction,
why does one observe a bigger effective dynamic
exponent $z$ but a correct static exponent $\eta$?
Qualitatively, indeed the logarithmic corrections
in both $M^{(2)}(t)$ and $A(t)$ effectively result in
a bigger $z$. But it is probably only by chance
that a correct $\eta$ is quantitatively kept. 

In conclusions, with Monte Carlo simulations
we have investigated the short-time behavior
of the dynamic processes starting from both
ordered and disordered states for the two-dimensional 
XY model. The results 
confirm that there is a logarithmic correction
to scaling in case of starting from a disordered state,
but an inverse power law correction
in case of starting from an ordered state.
The dynamic exponent is $z=2.04(1)$,
slightly bigger than the theoretical value, $z=2$.
We are satisfied with this result since for many
statistical systems $z$ is also different from the
'classical' value $z=2$.

{\bf Acknowledgements}:
This work is supported in part by NNSF of China and
DFG; TR 300/3-1.


\begin{table}[h]\centering 
\begin{tabular}{lllllllll}
 $d/z$ & $z_1$ & $\eta/2 z$ & $\eta$ &  $(2-\eta)/z$ & $z_2$  & $\theta$   
     & $d/z-\theta$  & $z_3$ \\ 
 \hline 
 0.982(10) & 2.04(2) &  0.0588(3) & 0.240(3)  &  0.866(3) & 2.03(1)
     &  0.250(2)   &  0.730(1)  &  2.04(1) 
 \end{tabular} 
\caption{Critical exponents measured for different dynamic processes.
The dynamic exponent $z_1$ is estimated from $d/z$. With $z_1$
as input, from $\eta/2 z$ we obtain $\eta$. With $\eta$ in hand,
 $z_2$ is calculated from $(2-\eta)/z$. From $\theta$ and $d/z-\theta$
 we estimate $z_3$.} 
\label{t1} 
\end{table}

\end{document}